\documentclass[aps,prl,reprint,groupedaddress,floatfix]{revtex4-1}
\usepackage[pdftex]{graphicx} 
\usepackage{epstopdf}
\usepackage{amssymb} 
\usepackage{amsmath}

\let\baraccent=\= 
\renewcommand{\=}[1]{\stackrel{#1}{=}} 

\begin{document}

\title{Photocurrent measurements of supercollision cooling in graphene }
\author{Matt W. Graham,$^{1,3}$ Su-Fei Shi,$^{1,3}$ Daniel C. Ralph,$^{1,3}$ Jiwoong Park$^{2,3}$  and Paul L. McEuen$^{1,3}$ }
\affiliation{$^{1.}$ Laboratory for Atomic and Solid State Physics, Cornell University, Ithaca, NY 14853, USA}
\affiliation{$^{2.}$ Department of Chemistry and Chemical Biology, Cornell University}
\affiliation{$^{3.}$ Kavli Institute at Cornell for Nanoscale Science, Ithaca, NY 14853, USA}
\begin{abstract}

The cooling of hot electrons in graphene is the critical process underlying the operation of exciting new graphene-based optoelectronic and plasmonic devices, but the nature of this cooling is controversial. We extract the hot electron cooling rate near the Fermi level by using graphene as novel photothermal thermometer that measures the electron temperature ($T(t)$) as it cools dynamically. We find the photocurrent generated from graphene $p-n$ junctions is well described by the energy dissipation rate $C dT/dt=-A(T^3-T_l^3)$, where the heat capacity is $C=\alpha T$ and $T_l$ is the base lattice temperature. These results are in disagreement with predictions of electron-phonon emission in a disorder-free graphene system, but in excellent quantitative agreement with recent predictions of a disorder-enhanced supercollision (SC) cooling mechanism. We find that the SC model provides a complete and unified picture of energy loss near the Fermi level over the wide range of electronic (15 to $\sim$3000 K) and lattice (10 to 295 K) temperatures investigated.  
\end{abstract}
\maketitle
How does an excited electron lose its energy? This is a central problem in fields ranging from condensed matter to particle physics. One key pathway is the emission of massless bosons such as photons or phonons. However, momentum must be conserved and the phase space available for such emissions can be dramatically restricted. For example, an electron moving through free space cannot emit a photon without transferring momentum to a third body.


\begin{figure}[htbp]
\includegraphics[height=2.7in]{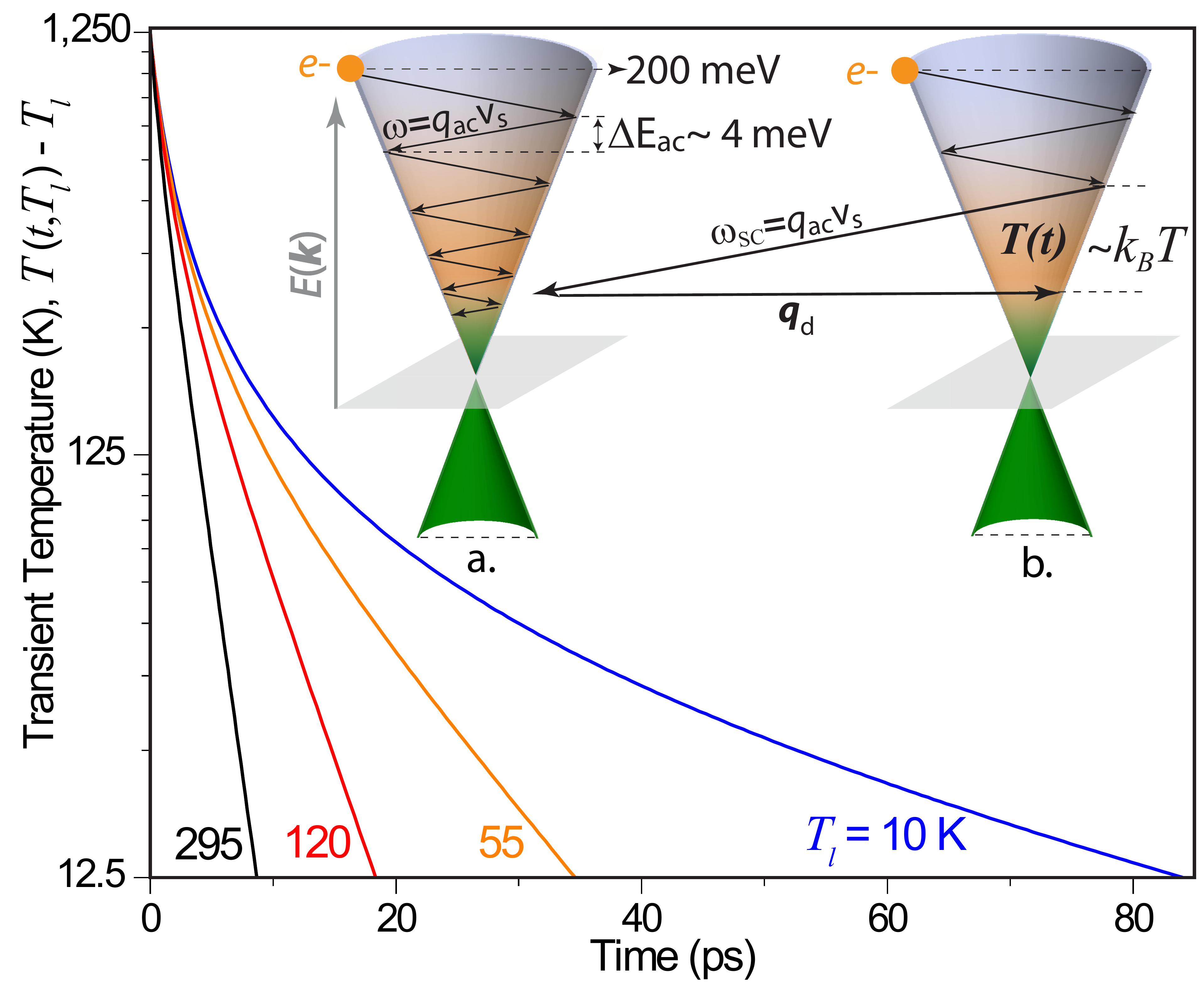}   
\caption{ \textbf{Hot electron cooling by acoustic phonons. (a)} Momentum conservation restricts cooling of hot carriers near the Fermi level (green) to low energy ($\lesssim 4$ meV) acoustic phonon emission (black arrows, scale exaggerated). \textbf{(b)} In a supercollision (SC) transition the momentum restrictions is relaxed by the lattice disorder ($\textbf{\textit{q}}_d$), enabling faster cooling by emission of high energy ($\sim k_BT$) acoustic phonons. Solving the SC rate law $dT/dt=-(A/\alpha)(T^3-T_l^3)/T$, we plot the predicted cooling of the graphene hot electron temperature $T(t,T_l)-T_l$ (log-scale).  The thermal decay changes from inverse to exponential with increasing lattice temperature.   
}
\end{figure}

\begin{figure*}[htbp]
\includegraphics[height=3.5in]{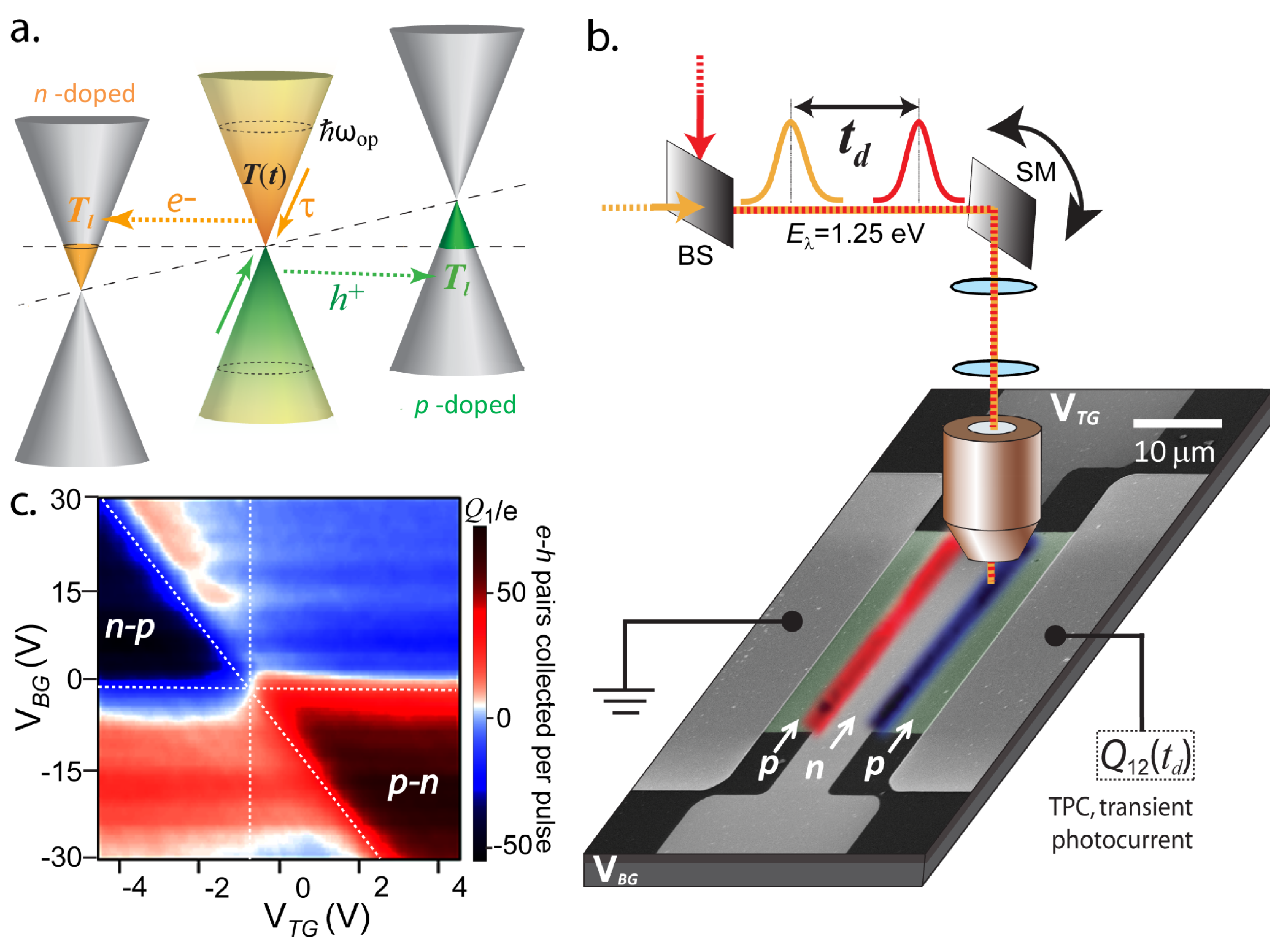}   
\caption{  \textbf{Photocurret setup, a time-resolved graphene thermometer. (a)} The temperature $T(t)$ of thermalized hot electrons and holes in a graphene $p-n$ junction cool at a characteristic rate $\tau^{-1}$. The elevated junction temperature drives the collected thermoelectric current, $i(t)$.  \textbf{(b)} Experimental setup where we collect PC from graphene as a function of $T_l$, laser power ($P$) and two-pulse time delay ($t_d$). Graphene (green, false color) is excited using a focused CW or pulsed laser. The device SEM shows an overlay of a spatial PC map with peaks for the graphene $p-n$ (0.7 nA, red) and $n-p$ junctions (-0.6 nA, blue). \textbf{(c)} Pulsed excitation PC map, plotting electrons collected ($Q_1/e$) vs. applied gate voltages. Tuning the electrostatic gates ($V_{TG}$, and $V_{BG}$) show six PC regions (dotted lines). 
  }
\end{figure*}

The cooling of hot electrons in graphene presents another interesting case. Here, hot electrons ($e^-$) move at a constant speed $v_F$ on a conical energy-momentum surface, and dissipate heat by phonon emission.  The optic phonon energies in graphene are unusually high, $\hbar \omega_{op} \gtrsim 200$ meV and mediate cooling for only very hot electrons.\cite{viljas2010} For electrons with energy below $\hbar \omega_{op}$, acoustic phonon emission is the dominant cooling pathway. However, these phonons move with the much slower sound velocity $v_s \ll v_F$. \cite{Tse2009,Bistritzer2009} As shown in Fig 1a, this velocity mismatch, combined with momentum conservation, greatly restricts the energy ($E_{ac}$) of emitted acoustic phonons to $\Delta E_{ac}/k_B T \leq 2v_s/v_F \sim$ 0.04 , where $k_BT$ is the typical energy of a hot electron.  More than $\sim$40 acoustic phonons to cool a hot electron to just half an initial energy of 0.2 eV.\cite{Kubakaddi2009,Tse2009}  This inefficient process creates a cooling bottleneck, with calculated cooling times exceeding $300$ ps.\cite{Bistritzer2009,Malic2011}  

Alternatively, a recent theory by Song \textit{et al.} predicts that disorder effectively relaxes the momentum conservation constraint, enabling the emission of large energy ($k_BT$) and momentum ($k_BT/\hbar v_s$) acoustic phonons, as shown in Fig. 1b. This mechanism is called supercollision (SC) cooling, and the theory predicts relaxation times of 1-10 ps, orders of magnitude faster than the disorder-free model.\cite{Song2011a} 

Here, we perform the first experiments to directly test the conflicting predictions of  hot electron models.\cite{Kubakaddi2009,Bistritzer2009,Song2011a,viljas2010,Sun2011} We use the photothermal effect in graphene as a novel quantitative probe of hot carrier cooling near the Fermi level. We find excellent agreement with the predictions of the supercollision model, showing that disorder effectively relaxes momentum conservation and leads to very rapid electron cooling. Using the cooling rates extracted, we directly determine the hot electron temperature in graphene, which is of central importance both to graphene's fundamental physics and for its use in a variety of electronic and optoelectronic applications such as photodetectors and bolometers.\cite{Mueller2010,Lemme2011,Bonaccorso2010,yan2012} 

The energy relaxation rate of a hot electron gas, $\frac{dE}{dt}= C \frac{dT}{dt}=P_{in}-H$ is determined by the heat loss rate $H$ and the heat capacity $C$,\cite{Bistritzer2009} where $P_{in}$ is the incident power delivered to the electrons. For a degenerate electron gas with heat capacity $C=\alpha T$ and $T > \hbar v_s \frac{k_F}{k_B}$ (typically only 5-10 K), the SC mechanism shown in Fig. 1b predicts $H_{SC}=A(T^3-T_l^3)$, where $A$ is rate coefficient, $k_F$ is the Fermi momentum and $T_l$ is the lattice temperature.\cite{Song2011a} For comparison, the conventional momentum conserving model (Fig. 1a) gives $H=A'T^4(T-T_l)$ for $E_F \ll k_B T$ or $H=A''(T-T_l)$ for $E_F \gg k_B T$, where $E_F$ is the Fermi energy.\cite{Bistritzer2009,viljas2010,Tse2009}

Under steady-state conditions where $H=P_{in}$, the SC model predicts the following temperature scaling with input power:  
\begin{align}
T & = (P_{in}/A)^{1/3},     \quad        & T \gg T_l \notag \\
T & = T_l + \frac{P_{in}}{3AT_l^2},  \quad   & T-T_l \ll T_l.
\end{align}
If instead we deliver a short impulse of energy $P_{in}=F_{in} \delta(t)$ to the system, the electron gas is heated to an initial temperature of $T_o=\sqrt{T_l^2+2F_{in}/\alpha}$, where $F_{in}$ is the remaining deposited energy after the initial optical phonon heat dissipation.\cite{breusing2009} The subsequent decay of transient electron gas temperature $T(t)$ is governed by $dT/dt=-H_{SC}/C$ with solutions:
\begin{align}
T(t) & = \frac{T_o}{1+t/\tau_o}, \quad         & T(t) \gg T_l \notag \\
T(t) & = T_l+(T_o-T_l)e^{\frac{-t}{\tau_1}},   \quad  & T(t)-T_l \ll T_l
\end{align}
where $\tau_o^{-1}=(A/\alpha)T_o$ and $\tau_1^{-1}=(3A/\alpha)T_l$ are characteristic hot electron cooling rates. The full solution for $T(t,T_l)$ is plotted in Fig. 1 using a rate coefficient $A/\alpha$ we later determine as $5.5 \times 10^8$ K$^{-1} s^{-1}$. With increasing $T_l$, the thermal cooling, $T(t,T_l)l$ changes from inverse to exponential in time. 

To experimentally test the above predictions, we locally heat a graphene $p-n$ junction with laser and use photocurrent generated as a thermometer of either the steady-state ($T_{CW}$) or transient ($T(t)$) hot electron temperatures. When heating graphene using 180 fs-long light pulses, we assume that only a fraction $\gamma=F_{in}/F$ of the total incident laser pulse energy ($F$) is retained in the hot electron gas created. This thermalized distribution is characterized by an initial temperature $T_o$, and cools dynamically at a rate $\tau^{-1}$ (see Fig. 2a).  Similarly under continuous wave (CW) illumination, only a fraction $\gamma = P_{in}/P$ of the total incident laser power $P$ is coupled into the electron gas, maintaining a steady-state temperature.    
 
In Fig. 2b, we show a schematic of the photocurrent measurement setup and single-layer graphene $p-n$ junction photodetector device. The junctions are created by globally $p$ (or $n$) doping the graphene sheet with an electrostatic back gate ($BG$), and locally $n$ (or $p$) doping through a top gate ($TG$).\cite{Lemme2011,Meric2008} We further overlay a spatial PC map on our device SEM image: Positive (red) or negative (blue) PC peaks are measured as we raster scan a 1.5 $\mu$m diameter laser spot over the $p-n$ and $n-p$ junctions, labeled. Data is collected at $T_l$=10 K unless otherwise indicated.  

\begin{figure}[htbp]
\includegraphics[height=7in]{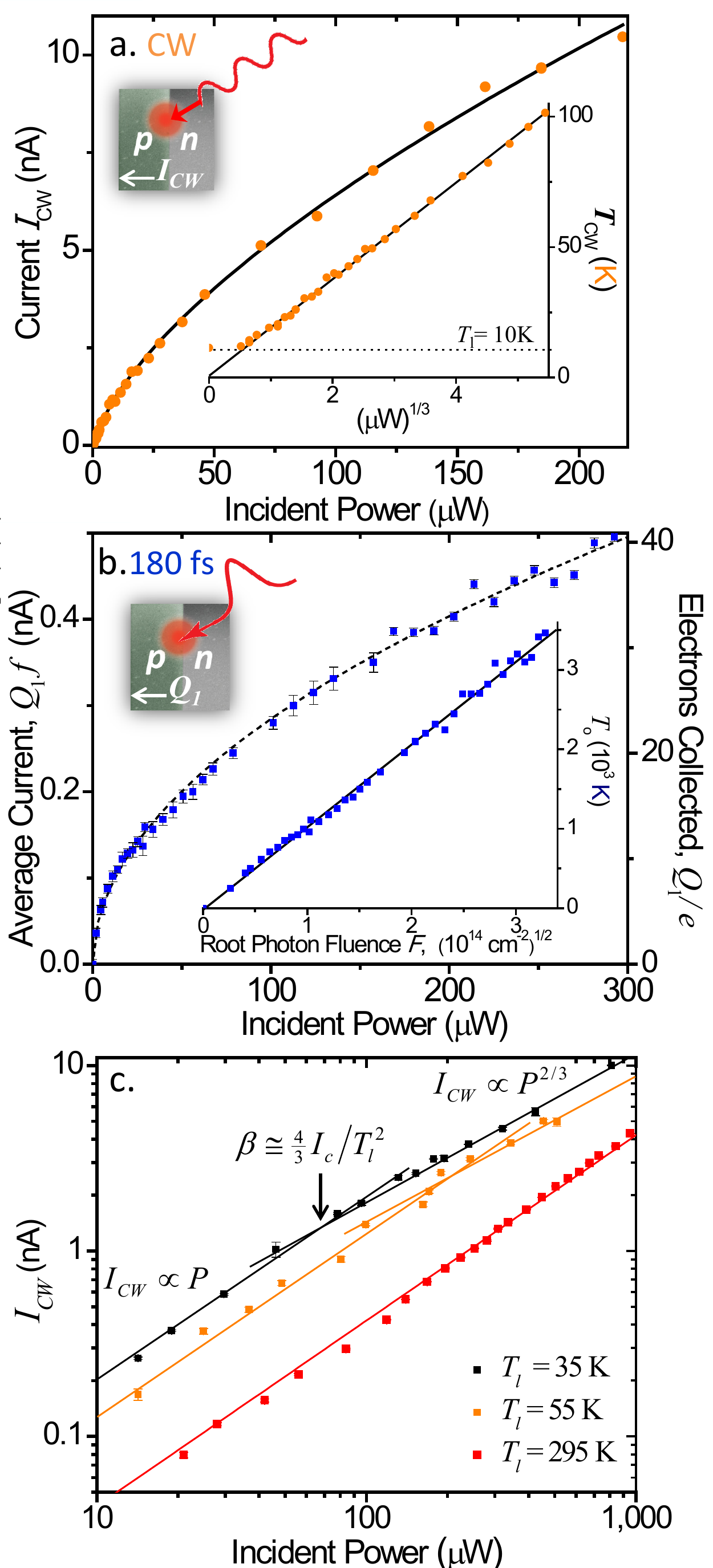}   
\caption{\textbf{PC response obeys SC power laws. (a)} PC generated under CW excitation at a graphene $p-n$ junction scales as $P^{0.65\pm0.02}$ (black line). (\textbf{\textit{inset}}) Corresponding electron temperatures scale linearly with $\sqrt[3]{P}$. \textbf{(b)} PC ($Q_1f$) and corresponding initial temperatures ($T_{o}$) vs. pulsed laser power. Black line, power law fit of $P^{0.50\pm0.03}$.  (\textbf{\textit{inset}}) Same plot converted to electrons collected per pulse ($Q_1/e$) vs. photon fluence (square root scale).\textbf{(c)} CW power dependence from $n-p$ junction becomes increasingly linear over the $T_l=$ 10-295 K range.  We estimate the cross-over current $I_c$, and use it to determine $\beta$.
}
\label{fig1}
\end{figure} 

Figure 2c plots the charge ($Q_1$) collected per excitation pulse using a laser repetition rate ($f$) of 76.1 MHz. Each pulse induces a time-dependent photocurrent response $i(t)$. We measure the resulting integrated charge $Q_1=\int{i(t)dt}$, or the average PC given by $Q_1f$. As the applied gate potentials are tuned, a sixfold pattern of alternating-sign photocurrent emerges, corresponding to $p-n$, $p-p^+$, $p^+-p$, $n-p$, $n-n^+$, and $n^+-n$ junctions. A similar pattern in also observed under CW excitation (see supplementary materials). It was recently shown in graphene that such six-fold PC patterns indicate electron-hole separation occurring by a thermoelectric process (illustrated in Fig. 2\textit{a}).\cite{Gabor2011, song2011}).  We can therefore use the measured thermoelectric current given by\cite{song2011,Xu2009}:   
\begin{equation}
i(t)=\beta T(t)(T(t)-T_l).
\end{equation} to extract the hot electron temperature for both CW and pulsed excitation conditions. Here $\beta$ is proportional to the Seebeck coefficient and is theoretically predicted to be $\sim 2$ pA/K$^2$ (see supplementary information). 

In Fig. 3a, we plot the PC collected at a $p-n$ junction ($V_{TG}$=2 V, $V_{BG}$=-15 V) as a function of CW laser power. The photocurrent is sublinear and is accurately fit by a power law, growing as $I_{CW} \sim P^{0.65 \pm 0.02}$. Figure 3b show an identical measurement using pulsed excitation. Comparing the two excitation techniques at identical laser powers, the amplitude of PC generated under pulsed excitation is at least 10 times smaller than the CW case, as was reported previously.\cite{Sun2012} Similar to the CW photocurrent, we find the current grows with a power law, but this time with: $I_{1p} \propto P^{0.50 \pm0.03}$ (see Fig. 3b \textit{inset}).  


\begin{figure}[htbp]
\includegraphics[height=7in]{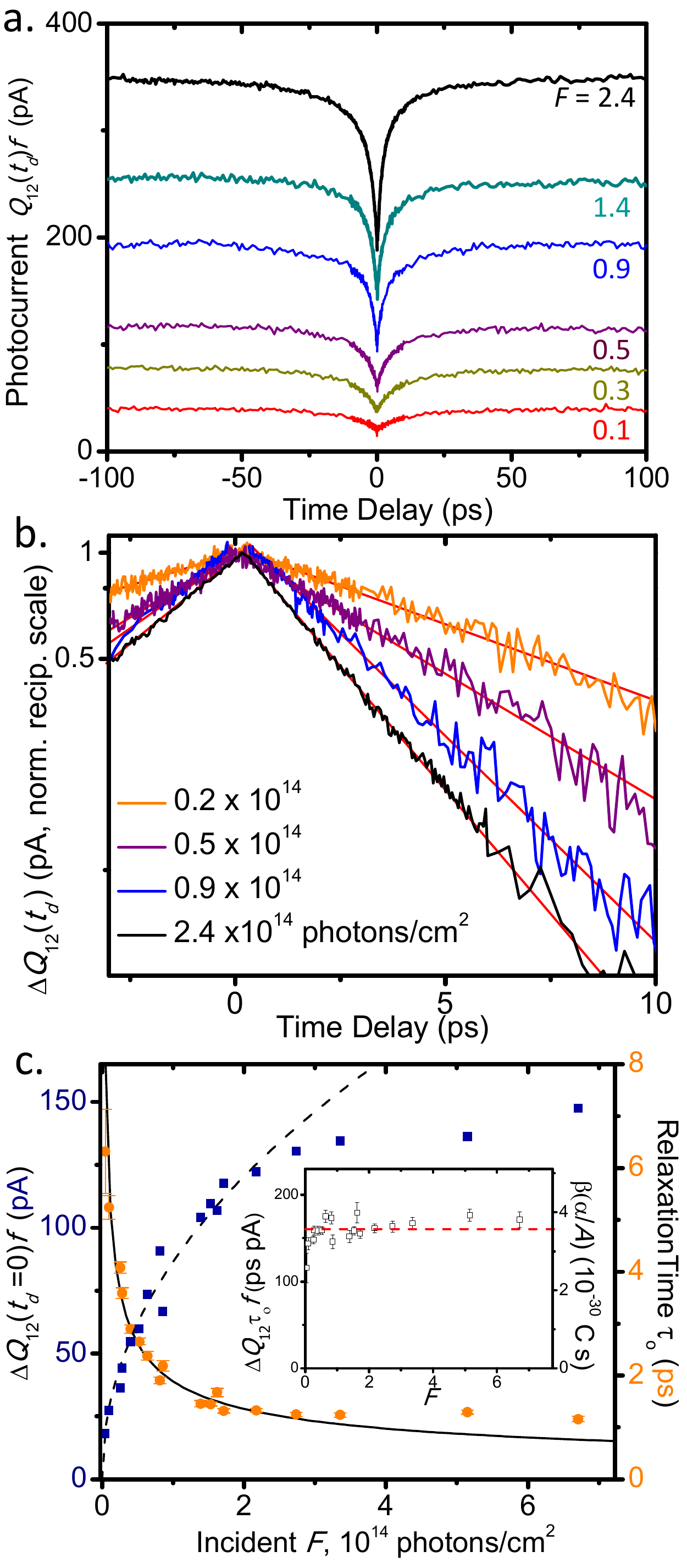} 
\caption{ \textbf{Extracting the hot electron relaxation time.} \textbf{(a)} Collected two-pulse photocurrent $(Q_{12}(t_d)f)$ response at selected incident photon fluences (in units of $\times 10^{14}$ photons/cm$^2$) \textbf{(b)} As shown, the decay of the TPC signal ($\Delta Q_{12}(t_d)f$, normalized) is closely linearized when plotted on an inverse scale. \textbf{(c)} TPC peak amplitude (square root fit, dashed line) and $\tau_o$ (inverse root fit, black line) vs. laser fluence, $F$. \textit{\textbf{(inset)}} The product of the data points yields a constant $\Delta Q_{12}(0) f \tau_o=160 \pm 13$ ps$\cdot$pA  (dashed red line) or an electron cooling rate of $A/\alpha$ of $\sim 5.5 \times 10^{8} K^{-1} s^{-1}$. }
\label{fig4}
\end{figure}

To compare the above power laws extracted against the SC model we combine equation (1) with the thermoelectric model to predict a CW PC power dependence of:
\begin{align}
I_{CW} &=\beta T_{CW}^2=\beta (P_{in}/A)^{2/3}  \quad      & T \gg T_l \notag \\
I_{CW} & \cong \frac {\beta P_{in}}{3AT_l}    \quad   & T-T_l \ll T_l.
\end{align} The fitted power laws are $I_{CW}\propto P^{0.65\pm0.02}$ for $T_l=10$ K and $I_{CW}\propto P$ for $T_l=295$ K (Fig. 3c), in excellent accord with the SC model.  Loss rates associated with other proposed momentum-conserving models $H \propto T$ or $H \propto T^5$ predict powers that are well outside of the error bars of the measured exponent.\cite{Bistritzer2009}  

The pulsed excitation power dependence can also be predicted using the SC model temperature $T(t)$ from equation (2):
	\begin{equation}
Q_1=\int^{\infty}_0{i(t)dt}=\beta(\alpha/A)T_o.
\end{equation}  The total current $Q_1 f$ collected is thus linearly related to the initial hot electron temperature, which from above is $T_o \cong \sqrt{2\gamma F/\alpha}$.  Hence the resulting PC should scale as $Q_1 \propto \sqrt{F}$, in excellent agreement with the data fits shown in Fig. 3b. 

\par The coefficient $\beta$ can also be extracted from Fig. 3 by finding the CW cross-over current ($I_c$) where the $I_{CW}$ power dependence transitions from $\sim P^{\frac{2}{3}}$ to linear in $P$. We find $\beta \cong \frac{4}{3} I_c/T_l^2$ (see supplementary materials).  Figure 3c shows $I_c$ occurs at higher powers as the base temperature is warmed to 295 K. We read-off cross-over currents of $\sim$85 pA ($T_l=$10 K), 1.2 nA (35 K) and 2.4 nA (55 K) and calculate a mean $\beta$ of $1.1$ pA/K$^2$ for our device.  In Fig. 3a (\textit{inset}) we use this $\beta$ to plot the graphene $p-n$ junction temperature vs. incident power. These direct measurements of the graphene electron temperature are important for the design and feasibility of graphene device exploiting electron thermal gradients.\cite{Mueller2010,Lemme2011,yan2012}
\\
\\
\indent We have shown the SC model coupled with the thermoelectric effect predicts the functional form of the CW and pulsed PC measurements. However, these PC power trends do not directly measure the timescales for electron cooling, nor the associated hot electron cooling rate $A/\alpha$ needed to quantitatively compare to the SC model and determine the absolute hot electron temperature in graphene.

	We use a time-dependent two-pulse excitation technique to measure $T(t)$ in graphene and extract the cooling rate.  The experimental setup is outlined schematically in Fig. 2b. The first pulse creates high energy $e-h$ pairs at the graphene $p-n$ junction, which rapidly thermalize and cool  to a temperature $T_o$ on a rapid $\lesssim 300$ fs timescale associated with optic phonon emission.\cite{breusing2009,Wang2010} The resulting distribution of hot electrons cools from $T_o$ to a transient temperature $T(t_d)$ by acoustic phonons at a characteristic rate $\tau_o^{-1}$.  At the pulse delay time $t_d$, a second collinear pulse of equal intensity is absorbed, heating the electron gas to $\sqrt{T_o +T(t_d)}$. The resulting total charge $Q_{12}(t_d)$ collected will then  vary with time-delay as the transient $p-n$ temperature ($T(t_d)$) cools.
	
In Fig. 4a the collected transient photocurrent(TPC) signal, $Q_{12}(t_d)f$ is plotted for selected photon fluences $F$. As $t_d \rightarrow 0$ ps, the magnitude of PC collected is greatly diminished because of the sublinear dependence of the PC on laser power (Fig. 3b). Analogous time-dependent reductions in PC have recently been recently reported for graphene based devices.\cite{Sun2011,Urich} In Fig. 4b, we plot $\Delta Q_{12}(t_d)$ on a normalized reciprocal scale.  The TPC decay kinetics are \textit{not} exponential, but instead show a striking resemblance to the $1/t_d$ thermal decay predicted by the SC model in equation (2). 

 To quantitatively interpret these results we integrate the time-dependent photothermal effect from equation (5) piecewise about $t_d$, giving: $Q_{12}(t_d)=\int^{t_d}_0{i(t,T_o)dt}+\int_{t_d}^{\infty}{i(t-t_d,\sqrt{T_o^2+T(t_d)^2})dt}$, where $T_o$ is the initial temperature created by each pulse independently. Solving using the thermal decay in equation (2) we obtain:
\begin{equation} 
\Delta Q_{12}(t_d) = \beta(\alpha/A) \left(T_o + T(t_d)- \sqrt{T_o^2+T(t_d)^2} \right). 
\end{equation}  This resulting TPC response function for $Q_{12}(t_d)$ is proportional to the transient temperature $T(t_d)$. Fig. 4b (red lines) shows this analytic solution fits our data well, requiring only two parameters; the amplitude $Q_1=\beta(\alpha/A)T_o$ and the thermal decay rate, $\tau_o^{-1}=AT_o/\alpha$.  This functional form further fits our data well for a wide range excitation wavelengths investigated (0.8 to 1.55 eV, data not shown).  

The TPC decay in Figures 4a and b becomes markedly faster with increasing photon fluence $F$.  This strong fluence dependence is captured by the SC model thermal decay in equation (2), which states $\tau_o= \alpha/A T_o^{-1}$, or equivalently that $\tau_o$ scales with $1/ \sqrt{F}$. Plotting the extracted fit parameters in Fig. 4c, we show hot electron cooling time ($\tau_o$, orange circles) decrease from 6.3 to 1.3 ps, closely scaling with $1/ \sqrt{F}$ (solid line fit), as predicted. The transient amplitude $\Delta Q_{12}(t_d=0)$ also scales nonlinearly as $\sqrt{F}$ (dotted line in Fig. 4\textsl{d}) up to a maximum $F$ of $3 \times 10^{14}$ photons$\cdot$cm$^{-2}$ where the transient response saturates. Their product $\Delta Q_{12} f \tau_o$ shown in Fig. 4d (\textit{inset}) is approximately constant. 

In the SC model, $Q_1 \tau_o=\beta(\alpha/A)^2$ measures the fundamental cooling rate coefficient $A/ \alpha$.  Using $\beta$=1.1 pA/K$^2$ found earlier, we find a SC cooling rate of $A/\alpha$ = $(5.5 \pm 0.4) \times 10^8$  K$^{-1}$s$^{-1}$.  

\begin{figure}[htbp]
\includegraphics[height=2.8in]{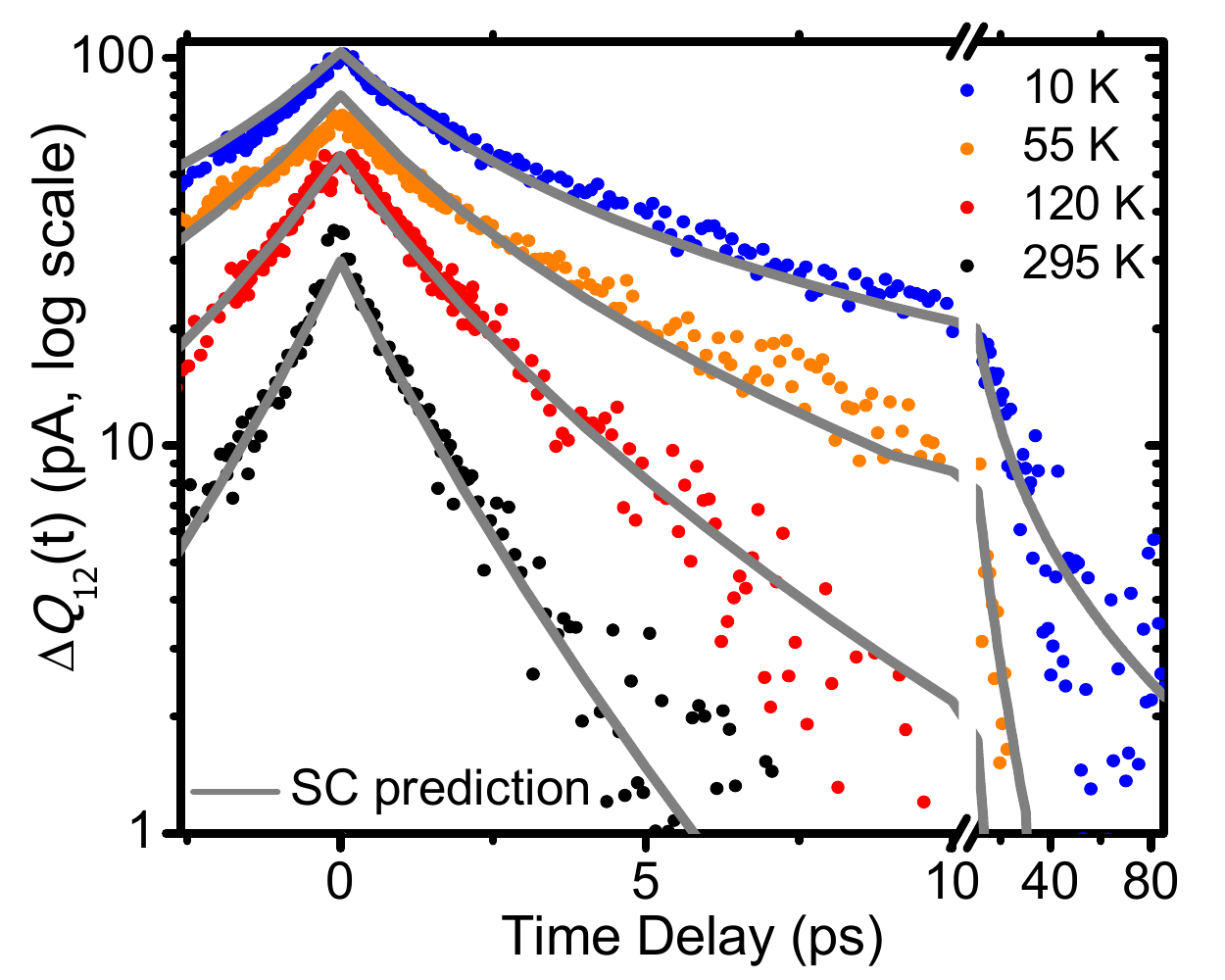}   
\caption{\textbf{SC model predicts TPC dependence on $T_l$.} At constant incident power, the TPC response varies considerably upon warming to 295 K. Using the SC temperature response in Fig. 1, we calculate the predicted TPC response $\Delta Q_{12}(t_d)$ with no free parameters (gray lines).
}
\end{figure}
Theoretical estimates of the SC cooling are given in Song $et$ $al.$ as\cite{Song2011a}: $\frac{A}{\alpha}=\frac{6 \zeta(3)}{\pi^2} \frac{\lambda}{k_F l} \frac{k_B}{\hbar} \cong \frac{2}{3}\frac{\lambda}{k_F l} \frac{k_B}{\hbar}$	
where the electron-phonon coupling strength is $\lambda=\frac{D^2}{\rho s^2} \frac{2E_F}{\pi(\hbar v_F)^2}$.\cite{Song2011a} Using estimates for the deformation potential, $D = 10-30$ eV, $E_F$=0.1 eV and a mean free path of $k_F l =10$, this theory predicts: 
$A/\alpha=1 \cdot 10^8 - 1 \times 10^{9}$ K$^{-1}$s$^{-1}$. (The range comes from the uncertainly in $D$). The best match to our experiments indicate $D = 12-18$ eV, well within the expected range. 

With our cooling rate coefficient now extracted, we now plot in Fig. 3b (\textit{inset}), the initial temperature, $T_o$ for the thermalized electron gas. $T_o$ can exceed 1000 K, an order magnitude higher than in the CW case.  Once heated, our data predicts hot electrons cool with a relaxation time varying inversely with $T_o$, as $\tau_o=((A/\alpha)T_o)^{-1}$ =1.8 ns/$T_o$[K].  

Since all the parameters in the model have been determined, the SC model predicts the lattice temperature dependence of the transient electron temperature with \textit{no free parameters}. Figure 5 plots TPC data for different base lattice temperatures for a constant laser photon fluence of $1.1 \times 10^{14}$ photons/cm$^2$, corresponding to $T_o \cong 1250$ K. Upon warming the lattice to room temperature, the amplitude of TPC signal shrinks by a factor of $\sim$3, and the kinetics exhibits a dramatic shift toward a rapidly decaying exponential function.  To compare with theory, we use the analytic SC model solutions for $T(t,T_l)-T_l$ plotted in Fig. 1 to numerically solve for the TPC response, $\Delta Q_{12}(t_d,T_l)$. With no adjustable parameters, the SC model curves in Fig. 5 accurately predicts both (i) the amplitudes, and (ii) strongly varying functional decay observed. Recent graphene time-resolved THz experiments report a similar change in decay kinetics with increasing $T_l$, as observed in Fig. 5 for our TPC data.\cite{Strait2011,Winnerl2011} 

The above results definitively show that the SC model gives an excellent quantitative description of both the CW and pulsed PC experiments. As a last demonstration of this connection, we connect the disparate magnitudes of the PC measured in the CW (Fig. 3a) and pulsed (Fig. 3b) excitation. The ratio, $I_{CW}/I_{1p}$ is predicted by the SC model using straightforward algebra to be $1/\sqrt[3]{4f\tau_o}$ or equivalently $\sim 1.2\sqrt[3]{T_o}$ (see supplementary information).  For the data shown in Fig. 3b, $T_o$ ranging from 250 K to 3500 K, giving corresponding $I_{CW}/I_{1p}$ ratio ranging from 8 to 18. This range is in excellent accord with the 10 to 20 range observed in Fig. 3, and provides an independent check that CW and pulsed PC experiments can be explained by the same fundamental underlying physics of SC hot electron cooling. 

In summary, we have introduced a quantitative framework for interpreting CW, one and two-pulse PC experiments as measurements of hot electron cooling of electrons near the Fermi energy. Over a broad range of electron (20-3000 K) and lattice (10-295 K) temperatures, we find the electron gas heat loss rate is $H_{SC}=A(T^3-T_l^3)$ with a rate coefficient $A/\alpha=5.5\times 10^8$ K$^{-1}$s$^{-1}$ for our device.  At low lattice temperature the associated cooling time given by $\tau_o = \left[(A/ \alpha)T_o \right]^{-1} = 1.8$ ns/$T_o$[K]. These cooling times are much faster than those predicted by acoustic phonon emission but are in excellent agreement with disorder-assisted supercollision cooling. The cooling rates extracted directly determine the graphene electron temperature, which is of central importance in designing graphene terahertz plasmonic devices, photodetectors and bolometers.

\section{Methods and Materials: }
\par Single-layer graphene on copper foil is grown using the Chemical Vapor Deposition (CVD) method \cite{Ruoff2009a}.  Micro-Raman was used to confirm the growth of large-grain single-layer graphene with no visible $D$ peak, indicative of high quality growth. Graphene was transferred using the lift-off technique onto a 300 nm SiO$_2$ layer grown on top of a silicon wafer which serves as the global back-gate ($BG$).  The large-grain growth graphene is divided into 30$\times$50 $\mu$m stripes using photolithography followed by oxygen plasma etching.  Electrode pads of titanium/gold(3nm/150nm) are deposited along graphene stripes with variable source-drain distances of 10 or 20 $\mu$m.  A good dielectric separation with the top gate is achieved with 10 nm of SiO$_2$ by electron beam deposition, followed by HfO$_2$ atomic layer deposition.  Finally, an optically translucent top gate of titanium/gold (2nm/20nm) is deposited along the center of the source-drain gap with a width of 6 $\mu$m.  The device is mounted in an Oxford HI-RES liquid helium cryostat. The CVD graphene photodetector device had a characteristic  high mobility of $\sim$8,000 cm$^2$V$^{-1}$s$^{-1}$ with centrally located Dirac points in conductance sweeps (see supplementary information). 

Light was generated by a Coherent MIRA oscillator that was externally compressed with via a prism-pair line. Autocorrelation measurements at the cryostat position yield beams centered at 1.25 eV and show a 180 fs FWHM pulse duration.  For TPC measurements the beam paths were cross-polarized to suppress pulse interference effects.  After a mechanical delay stage, the two beams are aligned in a collinear geometry at a beamsplitter (BS) and scanning mirror.  They are coupled into the microscope (Olympus BX-51) through a 50XIR Olympus objective with cover glass correction and piezo scanning mirror (SM, Fig. 2b). The TPC response is collected as function of time-delay ($t_d$) at 1 kHz beam modulation via lock-in and current amplifiers. 

\begin{acknowledgments}
\textbf{Acknowledgments}: this research was supported by the Kavli Institute at Cornell for Nanoscale Science (KIC), AFOSR (FA 9550-10-1-0410), by the NSF through the Center for Nanoscale Systems and by the MARCO Focused Research Center on Materials, Structures, and Devices. We thank Justin Song, Kathryn McGill and Josh Kevek for their helpful contributions.  Device fabrication was performed at the Cornell Nanofabrication Facility/National Nanofabrication Infrastructure Network.
\end{acknowledgments}



\begin{thebibliography}{10}

\bibitem{viljas2010}
J.~K. Viljas and T.~T. Heikkil$\ddot{\text{a}}$, {\em Physical Review B}, 2010, {\bf 81}(24),
  245404.

\bibitem{Tse2009}
W.-K. Tse and S.~{Das Sarma}, {\em Physical Review B}, 2009, {\bf 79}(23),
  235406.

\bibitem{Bistritzer2009}
R.~Bistritzer and A.~H. MacDonald, {\em Physical Review Letters}, 2009, {\bf
  102}(20), 206410.

\bibitem{Kubakaddi2009}
S.~S. Kubakaddi, {\em Physical Review B}, 2009, {\bf 79}(7), 075417.

\bibitem{Malic2011}
E.~Malic, T.~Winzer, E.~Bobkin, and A.~Knorr, {\em Physical Review B}, 2011,
  {\bf 84}(20), 205406.

\bibitem{Song2011a}
J.~C.~W. Song, M.~Y. Reizer, and L.~S. Levitov, {\em {arXiv:1111.4678}}, 2011.

\bibitem{Sun2011}
D.~Sun, C.~Divin, C.~Berger, W.~A. de~Heer, P.~N. First, and T.~B. Norris, {\em
  physica status solidi (c)}, 2011, {\bf 8}(4), 1194--1197.

\bibitem{Mueller2010}
T.~Mueller, F.~Xia, and P.~Avouris, {\em Nat Photon}, 2010, {\bf 4}(5),
  297--301.

\bibitem{Lemme2011}
M.~C. Lemme, F.~H.~L. Koppens, A.~L. Falk, M.~S. Rudner, H.~Park, L.~S.
  Levitov, and C.~M. Marcus, {\em Nano Lett.}, 2011, {\bf 11}(10), 4134--4137.

\bibitem{Bonaccorso2010}
F.~Bonaccorso, Z.~Sun, T.~Hasan, and A.~C. Ferrari, {\em Nat Photon}, 2010,
  {\bf 4}(9), 611--622.

\bibitem{yan2012}
J.~Yan, M.~Kim, J.~A. Elle, A.~B. Sushkov, G.~S. Jenkins, H.~M. Milchberg,
  M.~S. Fuhrer, and H.~D. Drew, {\em Nature Nanotechnology}, 2012.

\bibitem{breusing2009}
M.~Breusing, C.~Ropers, and T.~Elsaesser, {\em Physical Review Letters}, 2009,
  {\bf 102}(8), 086809.

\bibitem{Meric2008}
I.~Meric, M.~Y. Han, A.~F. Young, B.~Ozyilmaz, P.~Kim, and K.~L. Shepard, {\em
  Nature Nanotechnology}, 2008, {\bf 3}(11), 654--659.

\bibitem{Gabor2011}
N.~M. Gabor, J.~C.~W. Song, Q.~Ma, N.~L. Nair, T.~Taychatanapat, K.~Watanabe,
  T.~Taniguchi, L.~S. Levitov, and P.~Jarillo-Herrero, {\em Science}, 2011,
  {\bf 334}(6056), 648 --652.

\bibitem{song2011}
J.~C.~W. Song, M.~S. Rudner, C.~M. Marcus, and L.~S. Levitov, {\em Nano Lett.},
  2011, {\bf 11}(11), 4688--4692.

\bibitem{Xu2009}
X.~Xu, N.~M. Gabor, J.~S. Alden, A.~M. van~der Zande, and P.~L. McEuen, {\em
  Nano Lett.}, 2009, {\bf 10}(2), 562--566.

\bibitem{Sun2012}
D.~Sun, G.~Aivazian, A.~M. Jones, J.~S. Ross, W.~Yao, D.~Cobden, and X.~Xu,
  {\em Nat Nano}, 2012, {\bf 7}(2), 114--118.

\bibitem{Wang2010}
H.~Wang, J.~H. Strait, P.~A. George, S.~Shivaraman, V.~B. Shields,
  M.~Chandrashekhar, J.~Hwang, F.~Rana, M.~G. Spencer, C.~S. Ruiz-Vargas, and
  J.~Park, {\em Applied Physics Letters}, 2010, {\bf 96}, 081917.

\bibitem{Urich}
A.~Urich, K.~Unterrainer, and T.~Mueller, {\em Nano Letters}, 2011, {\bf 7}(1),
  2804.

\bibitem{Strait2011}
J.~H. Strait, H.~Wang, S.~Shivaraman, V.~Shields, M.~Spencer, and F.~Rana, {\em
  Nano Lett.}, 2011, {\bf 11}(11), 4902--4906.

\bibitem{Winnerl2011}
S.~Winnerl, M.~Orlita, P.~Plochocka, P.~Kossacki, M.~Potemski, T.~Winzer,
  E.~Malic, A.~Knorr, M.~Sprinkle, C.~Berger, W.~A. de~Heer, H.~Schneider, and
  M.~Helm, {\em Physical Review Letters}, 2011, {\bf 107}(23), 237401.

\bibitem{Ruoff2009a}
X.~Li, W.~Cai, J.~An, S.~Kim, J.~Nah, D.~Yang, R.~Piner, A.~Velamakanni,
  I.~Jung, E.~Tutuc, S.~K. Banerjee, L.~Colombo, and R.~S. Ruoff, {\em
  Science}, 2009, {\bf 324}(5932), 1312--1314.

\end{thebibliography}

\end{document}